\documentclass[12pt,prd,tightenlines,nofootinbib,showpacs]{revtex4}
\usepackage{bm}
\usepackage{graphics}
\usepackage{epsfig}
\usepackage{amsmath}
\usepackage{amsfonts}
\newcommand{\be}{\begin{equation}}
\newcommand{\ee}{\end{equation}}
\newcommand{\bdis}{\begin{displaymath}}
\newcommand{\edis}{\end{displaymath}}
\begin{document}
\title{\begin{flushright}{\rm\normalsize SSU-HEP-12/08\\[5mm]}\end{flushright}
Hyperfine structure of excited state of muonic helium atom}
\author{\firstname{A.A.} \surname{Krutov}}
\email{aakrutov@samsu.ru}
\affiliation{Samara State University, Pavlov street 1, 443011, Samara, Russia}
\author{\firstname{A.P.} \surname{Martynenko}}
\email{a.p.martynenko@samsu.ru}
\affiliation{Samara State University, Pavlov Street 1, 443011, Samara, Russia}
\affiliation{Samara State Aerospace University named after S.P. Korolyov, Moskovskoye Shosse 34, 443086,
Samara, Russia}

\begin{abstract}
The recoil, vacuum polarization and electron vertex corrections
of first and second orders in the fine structure constant $\alpha$ and the
ratio of electron to muon and electron to $\alpha$-particle masses
are calculated in the hyperfine splitting of the $1s^{(e)}_{1/2}2s^{(\mu)}_{1/2}$
state of muonic helium atom $(\mu e ^4_2He)$ on the basis of a perturbation
theory. We obtain total result for the muonically excited state hyperfine
splitting $\Delta \nu^{hfs}=4295.66$~MHz which
improves previous calculations due to the account of new
corrections and more accurate treatment of the electron vertex contribution.
\end{abstract}

\pacs{31.30.Jv, 12.20.Ds, 32.10.Fn}

\maketitle

\section{Introduction}

The muonic helium atom $(\mu e ^4_2He)$ represents the simple three-body atomic
system. The interaction between magnetic moments of the muon and electron
leads to the hyperfine structure (HFS) of the $S$-wave energy levels.
Whereas the ground state hyperfine splitting was measured many years ago with
sufficiently high accuracy in \cite{Orth1,Orth2}
$\Delta\nu^{hfs}_{exp}$=4465.004(29)~MHz, the excited states in muonic helium atom
were studied significantly smaller. Nevertheless, during the formation of muonic helium
atoms the negative muon is kept by positive $\alpha$-particle and is found to be
in excited state. The lifetime of excited state with the muon locating in the $2S$-state is
sufficient to measure its hyperfine splitting. This can be important for the check of quantum
electrodynamics for three-particle muonic bound states.

At present, hydrogen-like bound-state physics in QED is reliably developed \cite{HBES,SY,kp1993,EGS,SGK}.
Contrary to two-particle bound states whose energy levels were
accurately calculated in quantum electrodynamics \cite{HBES,SY,kp1993,EGS,SGK}, the
hyperfine splitting of three-particle muonic states was studied
on the basis of perturbation theory (PT) and variational approach with smaller accuracy
\cite{LM,HH,AF,borie,kp2001}. There are several calculations devoted to HFS of the excited
state $1s^{(e)}_{1/2}2s^{(\mu)}_{1/2}$ in muonic helium atom \cite{RD,Amusia,Chen,barcza}.
In these papers different order corrections to HFS were studied by perturbation
theory, variational method and Born-Oppenheimer approach. The accuracy of the calculations
lies in the interval of several tenth parts of MHz. The variational method gives
high numerical accuracy of the calculation what was demonstrated in~\cite{RD,Chen,AF,kp2001}.
The correlated wave functions of the state $1s^{(e)}_{1/2}2s^{(\mu)}_{1/2}$ were constructed
and higher order corrections to the hyperfine splitting were also accounted for in this approach
in~\cite{Chen}.
It is appropriate to mention here that three-particle kinematics was chosen for the description
of muonic helium in~\cite{RD,Amusia,HH,LM,Chen} by different ways. So, numerical results of
obtained corrections of different order in $\alpha$ and the ratio of the particle masses are difficult
for direct comparison.

Bound particles in muonic helium have different masses $m_e\ll m_\mu\ll m_\alpha$. As a
result the muon and $\alpha$-particle compose the pseudonucleus $(\mu ^4_2He)^+$ and muonic
helium atom looks as a two particle system in first approximation. The existence of such mass hierarchy
enables to formulate a perturbation theory for the calculation of the energy levels.
The perturbation theory approach to the investigation of the hyperfine structure of muonic helium
based on nonrelativistic Schr\"odinger equation was developed in~\cite{LM}. Using their method we
describe the excited state in three-particle bound system $(\mu e ^4_2He)$ by the Hamiltonian:
\begin{equation}
\label{eq:ham}
H=H_0+\Delta H+\Delta H_{rec},~~~H_0=-\frac{1}{2M_\mu}\nabla^2_\mu-\frac{1}{2M_e}
\nabla^2_e-\frac{2\alpha}{x_\mu}-\frac{\alpha}{x_e},
\end{equation}
\begin{equation}
\label{eq:deltah}
\Delta H=\frac{\alpha}{x_{\mu e}}-\frac{\alpha}{x_e},~~~\Delta H_{rec}=-\frac{1}{m_\alpha}
{\mathstrut\bm\nabla}_\mu\cdot{\mathstrut\bm\nabla}_e,
\end{equation}
where ${\bf x_\mu}$ and ${\bf x_e}$ are the coordinates of the muon
and electron relative to the helium nucleus,
$M_e=m_em_\alpha/(m_e+m_\alpha)$, $M_\mu=m_\mu
m_\alpha/(m_\mu+m_\alpha)$ are the reduced masses of subsystems $(e
^4_2He)^+$ and $(\mu ^4_2He)^+$. The hyperfine part of the Hamiltonian
contributing to HFS of $S$-states is
\begin{equation}
\label{eq:hfsh}
\Delta H^{hfs}=-\frac{8\pi\alpha(1+\kappa_\mu)}{3m_em_\mu}
\frac{({\mathstrut\bm\sigma}_e{\mathstrut\bm\sigma}_\mu)}{4}\delta({\bf
x_\mu}-{\bf x_e}),
\end{equation}
where ${\mathstrut\bm \sigma}_e$ and ${\mathstrut\bm\sigma}_\mu$ are
the spin matrices of the electron and muon.
In initial approximation the wave function of the excited state $1s^{(e)}_{1/2}2s^{(\mu)}_{1/2}$ has the form:
\begin{equation}
\label{eq:wave}
\Psi_0({\bf x_e},{\bf x_\mu})=\psi_e({\bf x_e})\psi_\mu({\bf x_\mu})=\frac{1}{2\sqrt{2}\pi}
(W_\mu W_e)^{3/2}\left(1-\frac{W_\mu x_\mu}{2}\right)e^{-W_\mu x_\mu}e^{-W_e x_e},
\end{equation}
where $W_\mu=2\alpha M_\mu$, $W_e=\alpha M_e$.
Then basic contribution to the singlet-triplet hyperfine splitting can be calculated
analytically from contact interaction \eqref{eq:hfsh}:
\begin{equation}
\label{eq:ef}
\Delta\nu^{hfs}_0=<\frac{8\pi\alpha(1+\kappa_\mu)}{3m_em_\mu} \delta({\bf
x_\mu}-{\bf
x_e})>=\frac{\nu_F(1+\kappa_\mu)}{\left(1+\frac{M_e}{M_\mu}\right)^5}\left(1-\frac{M_e}{M_\mu}+
\frac{M_e^2}{M_\mu^2}\right), ~~~\nu_F=
\frac{8\alpha^4M_e^3}{3m_em_\mu}.
\end{equation}
Numerically, the Fermi splitting is equal $\nu_F=4516.915$ MHz. We
express further the hyperfine splitting contributions in the
frequency unit using the relation $\Delta E^{hfs}=2\pi\hbar\Delta
\nu^{hfs}$. For the sake of definiteness we write all numerical values
with an accuracy 0.001 MHz. The recoil correction of first order in the ratio
$M_e/M_\mu$ in~\eqref{eq:ef} amounts to
$\Delta\nu^{hfs}_{rec}=-6\frac{M_e}{M_\mu}\nu_F=-134.769$~MHz.
The muon anomalous magnetic moment correction is equal to $\nu_F \kappa_\mu$=5.266~MHz.
Modern numerical values of fundamental physical
constants are taken from the paper \cite{MT}: the electron mass
$m_e=0.510998928(11)\cdot 10^{-3}~GeV$, the muon mass
$m_\mu=0.1056583715(35)~GeV$, the fine structure constant
$\alpha^{-1}=137.035999074(44)$, the helium mass $m_\alpha$ =
3.727379240(82)~GeV, the electron anomalous magnetic moment
$\kappa_e= 1.15965218076(27)\cdot 10^{-3}$, the muon anomalous
magnetic moment $\kappa_\mu=1.16592091(63)\cdot 10^{-3}$.

In this work we aim to extend the approach of Lakdawala and Mohr \cite{LM}
to the case of HFS of the excited state $1s^{(e)}_{1/2}2s^{(\mu)}_{1/2}$ in
muonic helium atom. The terms of the Hamiltonian $\Delta H$ and $\Delta H_{rec}$
give important recoil corrections in second order perturbation theory.
A feature that distinguishes light muonic atoms among other simplest atoms is that
the structure of their energy levels depends strongly on the vacuum
polarization (VP) effects, nuclear structure and recoil corrections \cite{M1,M2,M3,udj}.
So, we investigate the role of all such potentially important corrections to excited state
HFS in muonic helium. Another purpose of our study consists in
the improved evaluation of the electron one-loop vertex corrections
to HFS of order $\alpha^5$ using analytical expression of the electron
electromagnetic form factors \cite{M4}.

\section{Recoil corrections}

Along with small parameter $\alpha$ there are two other small parameters
in this task equal to the ratio of the particle masses $M_e/M_\mu$ and $M_e/m_\alpha$.
They determine the recoil corrections which can be calculated by the perturbation theory.
First recoil correction is presented in~\eqref{eq:ef}. Let us consider the recoil contribution
in second order perturbation theory. For its study we use the initial expression for
second order correction to the HFS from~\cite{LM,M4}:
\be
\label{eq:basic}
\Delta\nu^{hfs,(2)}=\frac{16\pi\alpha(1+\kappa_\mu)}{3m_em_\mu}\int d{\bf x}_1\int d{\bf x}_2
\int d{\bf x}_3\psi_{\mu}^\ast({\bf x}_3)\psi_{e~0}^\ast({\bf x}_3)\times
\ee
\bdis
{\sum_{n,m}}'\frac{\psi_{\mu,n}({\bf x}_3)\psi_{e,m}({\bf x}_3)\psi^\ast_{\mu,n}({\bf x}_2)
\psi^\ast_{e,m}({\bf x}_1)}{E_{\mu}+E_e-E_{\mu,n}-E_{e,m}}\Delta H({\bf x}_2,{\bf x}_1)\psi_\mu({\bf x}_2)
\psi_e({\bf x}_1),
\edis
where the superscript $(2)$ designates the correction in second order perturbation theory.
The prime near the sum symbol means that the term with $n=2, m=0$ is excluded.
$\Delta H({\bf x}_1,{\bf x}_2)$ is determined by~\eqref{eq:deltah}, $E_\mu$ is the Coulomb muon energy
in the $2S$-state,
$E_e$ is the Coulomb electron energy in the $1S$-state.
It is useful to extract in~\eqref{eq:basic} several contributions. If the
muon is in the $2S$ intermediate state $(E_\mu-E_{\mu,n=2}=0)$ then
\be
\label{eq:sopt2s}
\Delta\nu^{hfs,(2)}_{2S}=\frac{16\pi\alpha(1+\kappa_\mu)}{3m_em_\mu}\int d{\bf x}_1\int d{\bf x}_3|\psi_{\mu}({\bf x}_3)|^2
\psi_e({\bf x}_3) \Delta V_\mu({\bf x}_1)\psi_e({\bf x}_1)G_e({\bf x}_1,{\bf x}_3),
\ee
where the Coulomb reduced Green's function of the electron (see \cite{Hameka})
\begin{equation}
\label{eq:green}
G_e({\bf x}_1,{\bf x}_3)=\sum_{n\not =0}^\infty\frac{\psi_{e,n}({\bf x}_3)\psi_{e,n}^\ast({\bf x}_1)}
{E_{e}-E_{e,n}}=-\frac{\alpha M_e^2}{\pi}e^{-W_e(x_1+x_3)}\Biggl[\frac{1}{2W_e x_>}-
\end{equation}
\begin{displaymath}
-\ln(2W_ex_>)-\ln(2W_ex_<)+Ei(2W_e x_<)+
\frac{7}{2}-2C-W_e(x_1+x_3)+\frac{1-e^{2W_e x_<}}{2W_e x_<}\Biggr],
\end{displaymath}
$x_<=min(x_1,x_3)$, $x_>=max(x_1,x_3)$, $C=0.577216\ldots$ is the Euler's constant and
$Ei(x)$ is the exponential-integral function,
\be
\label{eq:deltavmu}
\Delta V_\mu({\bf x}_1)=\int d{\bf x}_2\psi^\ast_\mu({\bf x}_2)\Delta H({\bf x}_1,{\bf x}_2)\psi_\mu({\bf x}_2)=
-\frac{\alpha}{x_1}e^{-W_\mu x_1}\left[1+\frac{3}{4}W_\mu x_1+\frac{1}{4}W_\mu^2x_1^2+\frac{1}{8}W_\mu^3x_1^3\right].
\ee
Substituting \eqref{eq:deltavmu} and \eqref{eq:green} in \eqref{eq:basic} we perform analytical integration
over all coordinates and expand exact result in the ratio $M_e/M_\mu$ preserving the terms of third order:
\be
\label{eq:nu_sopt_2s}
\Delta\nu^{hfs,(2)}_{2S}=\nu_F(1+\kappa_\mu)\frac{M_e}{M_\mu}\Biggl[\frac{631}{256}+\frac{M_e}{M_\mu}
\left(-14\ln\frac{M_e}{M_\mu}-\frac{5291}{512}-14\ln 2\right)+
\ee
\bdis
+\left(\frac{M_e}{M_\mu}\right)^2
\left(139\ln\frac{M_e}{M_\mu}-\frac{17361}{512}+139\ln 2\right)\Biggr]=61.127~MHz.
\edis

Another correction in~\eqref{eq:basic} is determined by the muon in the $2P$ intermediate state:
\be
\label{eq:nu_sopt_2p}
\Delta\nu^{hfs,(2)}_{2P}=-\frac{16\pi\alpha^2M_e^2(1+\kappa_\mu)W_\mu^5}{9m_em_\mu}\int x_2 d{\bf x}_2
\int x_3d{\bf x}_3\psi^\ast_\mu({\bf x}_3)\psi^\ast_e({\bf x}_3)\sum_{m=-1}^{1}Y_{1m}({\bf n}_3)Y^\ast_{1m}({\bf n}_2)\times
\ee
\bdis
\int d{\bf x}_1e^{-W_\mu x_2/2}e^{-W_\mu x_3/2}\Delta H({\bf x}_2,{\bf x}_1)\psi_\mu({\bf x}_2)\psi_e({\bf x}_1)
\sum_{m'=-1}^{1}Y_{1m'}({\bf n}_1)Y^\ast_{1m'}({\bf n}_3)g_{11}(W_ex_1,W_ex_3),
\edis
where the unit vectors ${\bf n}_i={\bf r}_i/r$ ($i=1,2,3$), the partial Green's function (see \cite{johnson})
\be
\label{eq:g11}
g_{11}(W_ex_1,W_ex_3)=e^{-W_e(x_1+x_3)}\Bigl(\frac{1}{4W_e^2x_>^2}+\frac{1}{2W_ex_>}+\frac{1}{2}\Bigr)
\Bigl(\frac{e^{2W_ex_<}}{4W_e^2x_<^2}-\frac{1}{4W_e^2x_<^2}-\frac{1}{2W_ex_<}-\frac{1}{2}\Bigr).
\ee
Integrating analytically in~\eqref{eq:nu_sopt_2p}, we can present the expansion of final result in $M_e/M_\mu$
up to terms of third order in the form:
\be
\label{eq:nu_sopt_2pa}
\Delta\nu^{hfs,(2)}_{2P}=\nu_F(1+\kappa_\mu)\left(-\frac{837}{256}\frac{M_e}{M_\mu}+\frac{12141}{512}\frac{M_e^2}{M_\mu^2}-
\frac{43335}{512}\frac{M_e^3}{M_\mu^3}\right)=-70.919~MHz.
\ee
In addition, there exists in~\eqref{eq:basic} the contribution corresponding to excited states of the muon:
\begin{equation}
\label{eq:muonn}
\Delta\nu^{hfs,(2)}_{n\not= 2S,2P}=-\frac{16\pi\alpha(1+\kappa_\mu)}{3m_em_\mu}\int d{\bf x}_1\int d{\bf x}_2\int d{\bf x}_3
\psi_\mu^\ast({\bf x}_3)\psi_e^\ast({\bf x}_3)\times
\ee
\bdis
\sum_{n\not= 2}\psi_{\mu,n}({\bf x}_3)\psi_{\mu,n}^\ast({\bf x}_2)G_e({\bf x}_3,{\bf x}_1,E_\mu+E_e-E_{\mu,n})
\frac{\alpha}{|{\bf x}_2-{\bf x}_1|}\psi_\mu({\bf x}_2)\psi_e({\bf x}_1).
\edis
It can be calculated analytically in the leading order in $M_e/M_\mu$ if we replace the exact Green's function
of the electron $G_e({\bf x}_3,{\bf x}_1,E_\mu+E_e-E_{\mu,n})$ by free Green's function
\be
\label{eq:free}
G_{e,0}({\bf x}_3,{\bf x}_1,E_\mu+E_e-E_{\mu,n})=\frac{M_e}{2\pi}\frac{e^{-b|{\bf x}_3-{\bf x}_1|}}
{|{\bf x}_3-{\bf x}_1|}, \qquad b=\sqrt{2M_e(E_{\mu,n}-E_\mu-E_e)}.
\ee
Replacing also the electron wave function by its value at the origin $\psi_e(0)$ and integrating over the
coordinate ${\bf x}_1$ in~\eqref{eq:muonn} we obtain (see \cite{LM,M4}):
\be
\label{eq:jj}
I_1=\int d{\bf x}_1\frac{e^{-b|{\bf x}_3-{\bf x}_1|}}{|{\bf x}_3-{\bf x}_1||{\bf x}_2-{\bf x}_1|}=
4\pi\left[\frac{1}{b}-\frac{1}{2}|{\bf x}_2-{\bf x}_3|+\frac{b}{6}|{\bf x}_2-{\bf x}_3|^2+...\right],
\ee
where the expansion in $b/W_\mu\sim \sqrt{M_e/M_\mu}$ is performed. The term $1/b$ does not contribute.
Taking second term in~\eqref{eq:jj}, we use further the completeness condition:
\be
\label{eq:delta}
\sum_{n\not=2P,2S}\psi_{\mu,n}({\bf x}_3)\psi^\ast_{\mu,n}({\bf x}_2)=\delta({\bf x}_3-{\bf x}_2)-
\psi_{\mu,2S}({\bf x}_3)\psi^\ast_{\mu,2S}({\bf x}_2)-\sum_{m=-1}^1\psi_{\mu,2P,m}({\bf x}_3)\psi^\ast_{\mu,2P,m}({\bf x}_2),
\ee
where we write explicitely subscripts $2S$ and $2P$ corresponding to $n=2$ muonic states.
Then the contribution of each component from~\eqref{eq:delta} can be calcuated analytically together with
the term $\frac{1}{2}|{\bf x}_2-{\bf x}_3|$.
We obtain following results corresponding to the $2S$ and $2P$ states in~\eqref{eq:delta}:
\be
\label{eq:2S2P}
\Delta\nu^{hfs,(2)}_{1}=-\nu_F(1+\kappa_\mu)\frac{M_e}{M_\mu}\frac{2167}{256},~~~
\Delta\nu^{hfs,(2)}_{2}=\nu_F(1+\kappa_\mu)\frac{M_e}{M_\mu}\frac{837}{256}.
\ee
Note that the order $O(M_e/M_\mu)$ corrections in~\eqref{eq:nu_sopt_2pa} and \eqref{eq:2S2P} corresponding
to $2P$-state cancel in the sum.
To increase the accuracy of the calculation we analyze third term in~\eqref{eq:jj}. It can be expressed
in terms of the following sum \cite{HBES,LM}:
\be
\label{eq:ss}
S_{1/2}=\sum_{n\not=2}\left(\frac{E_{\mu,n}-E_{\mu,2}}{R_\mu}\right)^{1/2}|<\mu,2S|\frac{{\bf x}}{a_\mu}|\mu,n>|^2.
\ee
The discrete and continuum states contributions to~\eqref{eq:ss} are equal correspondingly \cite{VAF}:
\be
\label{eq:sss}
S_{1/2}^{d}=\sum_{n>2}\frac{2^{16}n^6(n^2-1)(n-2)^{2n-6+\frac{1}{2}}}{(n+2)^{2n+6-\frac{1}{2}}}=4.918...,
\ee
\bdis
S_{1/2}^{c}=\int_0^\infty 2^5(k^2+\frac{1}{4})^{1/2}\frac{(k^2+1)}{(k^2+\frac{1}{4})^6}
\frac{k}{1-e^{-\frac{2\pi}{k}}}e^{-\frac{4}{k}arctg(2k)}=1.563...  .
\edis
As a result the contribution of third term from the expansion~\eqref{eq:jj} is equal to
\be
\label{eq:s12}
\Delta\nu^{hfs,(2)}_3=\nu_F(1+\kappa_\mu)\frac{2}{3}\left(\frac{M_e}{M_\mu}\right)^{3/2}S_{1/2}=6.853~MHz.
\ee
Another set of recoil corrections in second order PT is determined by the perturbation $\Delta H_{rec}=-\frac{1}{m_\alpha}\nabla_\mu\nabla_e$. In this case the basic contribution to HFS is
the following
\be
\label{eq:malpha}
\Delta\nu^{hfs}_\alpha=-\frac{32\pi\alpha^3M_eM_\mu}{3m_\alpha m_e m_\mu}\int d{\bf x}_1
\int d{\bf x}_2\int d{\bf x}_3\psi^\ast_{\mu}({\bf x}_3)\psi^\ast_e({\bf x}_3)\times
\ee
\bdis
\sum_{n,m}\frac{\psi_{\mu,n}({\bf x}_3)\psi_{e,m}({\bf x}_3)\psi^\ast_{\mu,n}({\bf x}_2)
\psi^\ast_{e,m}({\bf x}_1)}{E_{\mu}+E_e-E_{\mu,n}-E_{e,m}}\frac{W_e^{3/2}W_\mu^{3/2}}{2\pi\sqrt{2}}
e^{-W_ex_1}e^{-W_\mu x_2/2}\left(2-\frac{W_\mu x_2}{2}\right)({\bf n}_1{\bf n}_2),
\edis
where we substitute
\be
\label{eq:dd}
\nabla_2\nabla_1\psi_{\mu}({\bf x}_2)\psi_e({\bf x}_1)=\frac{\alpha^2 M_eM_\mu W_e^{3/2}
W_\mu^{3/2}}{2\pi\sqrt{2}}e^{-W_ex_1}e^{-W_\mu x_2/2}\left(2-\frac{W_\mu x_2}{2}\right)
({\bf n}_1{\bf n}_2).
\ee
The muon $2S$-state does not contribute in~\eqref{eq:malpha}. In the case of the $2P$ muon intermediate state
when $\psi_{\mu,n}$=$\psi_{\mu,2P}$, we can integrate at first over angle variables. Then we obtain
the radial integral $\int_0^\infty x_2^3 dx_2 e^{-x_2}(2-x_2/2)$=0. So, we should consider the muon
intermediate states with $n\not=2$ in~\eqref{eq:malpha}. Acting as in~\eqref{eq:muonn} we obtain:
\be
\label{eq:malpha1}
\Delta\nu^{hfs,(1)}_{\alpha}(n\not=2)=\frac{2\alpha^3M_e^2M_\mu W_e^3 W_\mu^3}{3m_\alpha m_e m_\mu \pi^2}
\int d{\bf x}_1\int d{\bf x}_2\int d{\bf x}_3(1-\frac{W_\mu x_3}{2})e^{-W_\mu x_3/2}e^{-W_\mu x_2/2}\times
\ee
\bdis
\left(2-\frac{W_\mu x_2}{2}\right)({\bf n}_1{\bf n}_2)\sum_{n\not=2}\psi_{\mu,n}({\bf x}_3)\psi^\ast_{\mu,n}({\bf x}_2)
\frac{e^{-b|{\bf x}_3-{\bf x}_1|}}{|{\bf x}_3-{\bf x}_1|}.
\edis
Then we can integrate over ${\bf x}_1$ and expand the result over small parameter $b$:
\be
\label{eq:jjj}
J_\alpha=\int d{\bf x}_1 ({\bf n}_1{\bf n}_2)\frac{e^{-b|{\bf x}_3-{\bf x}_1|}}{|{\bf x}_3-{\bf x}_1|}=
2\pi({\bf n}_2{\bf n}_3)\left[\frac{4x_3}{3b}-\frac{x_3^2}{2}+\frac{2bx_3^3}{15}+...\right].
\ee
Taking first term in square brackets we obtain two radial integrals which determine
the contribution~\eqref{eq:malpha1} for the transition $2S\to nP$ ($n>2$):
\be
\label{eq:j1j2}
J_1=\int_0^\infty e^{-x/2}e^{-x/n}\frac{2x}{n}L^3_{n-2}\left(\frac{2x}{n}\right)x^2dx, \quad
J_2=\int_0^\infty R_{2S}(x)R_{n1}(x)x^3dx.
\ee
Extracting two contributions to the product of $J_1$ and $J_2$ corresponding to descrete and
continuum states we present the correction~\eqref{eq:malpha1} in the form \cite{VAF}:
\be
\label{eq:malpha2}
\Delta\nu^{hfs,(1)}_{\alpha}(n\not=2)=\nu_F(1+\kappa_\mu)\frac{4M_e}{3m_\alpha}\sqrt{\frac{M_e}{M_\mu}}S_{j_1j_2}=
0.189~MHz,
\ee
\bdis
S_{j_1j_2}=
\sum_{n>2}\frac{2^{15}(n-2)^{2n-5-\frac{1}{2}}n^6(n^2-1)}{(n+2)^{2n+5+\frac{1}{2}}}
+\int_0^\infty\frac{2^{15} k (k^2+1)dk}{(4k^2+1)^{11/2}\left(1-e^{-\frac{2\pi}{k}}\right)}
e^{-\frac{4}{k}\arctan(2k)}=3.24101\ldots.
\edis
Second term $(-\pi({\bf n}_2{\bf n}_3)x_3^2)$ in the expansion~\eqref{eq:jjj} is not
dependent on $n$. So, we can present its contribution using the completeness condition:
\be
\label{eq:malpha2}
\Delta\nu^{hfs,(2)}_{\alpha}(n\not=2)=\frac{4\alpha^3M_e^2M_\mu W_e^3 W_\mu^3}{3\pi m_\alpha m_e m_\mu}
\int d{\bf x}_2\int x_3^2d{\bf x}_3\left(1-\frac{W_\mu x_3}{2}\right)e^{-W_\mu x_3/2}e^{-W_\mu x_2/2}\times
\ee
\bdis
\left(2-\frac{W_\mu x_2}{2}\right)({\bf n}_2{\bf n}_3)\left[\delta({\bf x}_2-{\bf x}_3)-
\psi_{\mu,2S}({\bf x}_3)\psi^\ast_{\mu,2S}({\bf x}_2)-\sum_{m=-1}^1\psi_{\mu,2P,m}({\bf x}_3)
\psi^\ast_{\mu,2P,m}({\bf x}_2)\right].
\edis
Only the $\delta$-term in~\eqref{eq:malpha2} gives nonzero correction of second order in the ratio of particle masses:
\be
\label{eq:malpha3}
\Delta\nu^{hfs,(2)}_{\alpha}(n\not=2)=\nu_F\frac{24M_e^2}{m_\alpha M_\mu}=0.074~MHz.
\ee

\begin{figure}[htbp]
\centering
\includegraphics[scale=0.8]{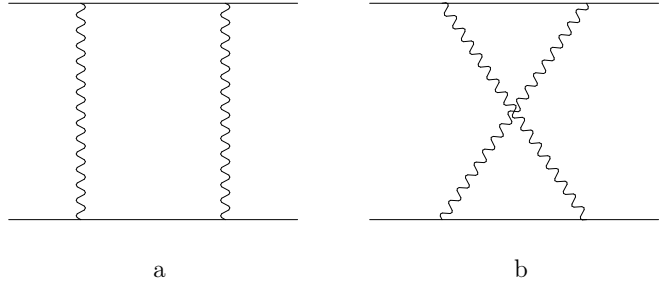}
\caption{Two photon exchange amplitudes in the electron-muon hyperfine
interaction.}
\label{fig4:diag4}
\end{figure}

Special attention has to be given to the recoil corrections
connected with two-photon exchange diagrams shown in Fig.\ref{fig4:diag4} in the
case of the electron-muon hyperfine interaction.  For the singlet-triplet
splitting the leading order recoil contribution to the interaction
operator between the muon and electron is determined as follows
\cite{EGS,Chen,RA}:
\be
\label{eq:2g}
\Delta V_{rec,\mu e,2\gamma}^{hfs}({\bf x}_{\mu e})=8\frac{\alpha^2}{m_\mu^2-m_e^2}\ln\frac{m_\mu}{m_e}
\delta({\bf x}_{\mu e}).
\end{equation}
Averaging the potential $\Delta V_{rec,\mu e,2\gamma}^{hfs}$ over the wave
functions~\eqref{eq:wave} we obtain the leading order recoil correction to the
hyperfine splitting:
\be
\label{eq:2g1}
\Delta\nu_{rec,\mu
e,2\gamma}^{hfs}=\nu_F\frac{3\alpha m_em_\mu}{\pi(m_\mu^2-m_e^2)}\ln\frac{m_\mu}
{m_e}=0.812~MHz.
\ee
There exist also the two-photon interactions between the bound particles of muonic
helium atom when one hyperfine photon transfers the interaction from the electron to
muon and another Coulomb photon from the electron to the nucleus (or from the muon to the nucleus).
Supposing that these amplitudes give smaller contribution to the hyperfine splitting we include
them in the theoretical error.

\section{Electron vertex corrections}

\begin{figure}[htbp]
\centering
\includegraphics[scale=0.8]{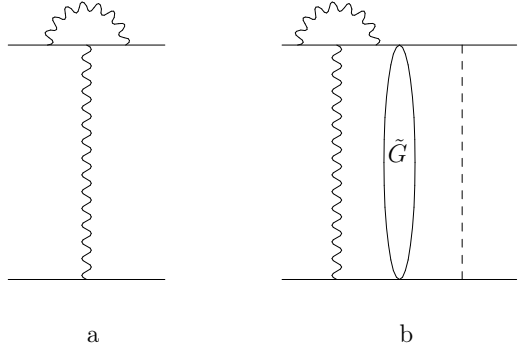}
\caption{The electron vertex corrections. The dashed line represents
the Coulomb photon. The wave line represents the hyperfine part of
the Breit potential. $\tilde G$ is the reduced Coulomb Green's
function.}
\label{fig1:diag1}
\end{figure}

In the initial approximation the potential of the hyperfine
splitting is determined by~\eqref{eq:hfsh}. It leads to the energy splitting
of order $\alpha^4$. In QED perturbation theory there is the
electron vertex correction to the potential~\eqref{eq:hfsh} which is defined by
the diagram in Fig.~\ref{fig1:diag1}(a). In momentum representation the
corresponding operator of hyperfine interaction has the form \cite{M4}:
\begin{equation}
\label{eq:amm1}
\Delta V^{hfs}_{vertex}(k^2)=-\frac{8\alpha^2(1+\kappa_\mu)}{3m_em_\mu}
\left(\frac{{\mathstrut\bm\sigma}_e{\mathstrut\bm\sigma}_\mu}{4}\right)
\left[G_M^{(e)}(k^2)-1\right],
\end{equation}
where $G_M^{(e)}(k^2)$ is the electron magnetic form factor. We
extracted for the convenience the factor $\alpha/\pi$ from
$\left[G_M^{(e)}(k^2)-1\right]$. Usually used approximation for the
electron magnetic form factor $G_M^{(e)}(k^2)\approx
G_M^{(e)}(0)=1+\kappa_e$ is not quite correct in this task. Indeed,
characteristic momentum of the exchanged photon is $k\sim\alpha
M_\mu$. It is impossible to neglect it in the magnetic form factor
as compared with the electron mass $m_e$. So, we should use exact
one-loop expression for the magnetic form factor
\cite{t4}. Trying to improve the previous estimation of the
correction due to the electron anomalous magnetic moment we will use
further exact one-loop expression for the Pauli form factor $g(k^2)$
known from the QED calculation (see \cite{t4}) setting $G_M^{(e)}(k^2)-1\approx g(k^2)$.

Using the Fourier transform of the potential~\eqref{eq:amm1} and averaging the
obtained expression over wave functions~\eqref{eq:wave} we represent the
electron vertex correction to the hyperfine splitting as follows \cite{M4}:
\begin{equation}
\label{eq:amm2}
\Delta\nu^{hfs}_{vert}=\nu_F\frac{\alpha(1+\kappa_\mu)M_e}{2\pi^2M_\mu}
\left(\frac{m_e}{\alpha M_\mu}\right)^3\int_0^\infty
\frac{k^2 g(k^2) dk(1-3k^2\frac{m_e^2}{W_\mu^2}+2k^4\frac{m_e^4}{W_\mu^4})}
{\left[1+k^2\left(\frac{m_e^2}{W_\mu^2}\right)\right]^4\left[\frac{M_e^2}{M_\mu^2}+
k^2\left(\frac{m_e^2}{W_\mu^2}\right)\right]^2}
=5.078~MHz.
\end{equation}
Let us remark that the contribution~\eqref{eq:amm2} is of order $\alpha^5$.
Numerical value~\eqref{eq:amm2} is obtained after numerical integration with
the exact one-loop expression of the Pauli electron form factor
$g(k^2)$ \cite{t4}. If we use the value $g(k^2=0)$ then the
electron vertex correction is equal $5.246$ MHz. So, using the exact
expression of the electron form factor $g(k^2)$ in the one-loop
approximation we observe small decrease of the vertex correction
to the hyperfine splitting from $1\gamma$ interaction. Taking the
expression~\eqref{eq:amm1} as an additional perturbation potential we have to
calculate its contribution to HFS in second order perturbation
theory (see the diagram in Fig.~\ref{fig1:diag1}(b)). In this case the dashed line
represents the Coulomb Hamiltonian $\Delta H$~\eqref{eq:deltah}. Following the
method of the calculation formulated in previous section
we divide again total contribution from the
amplitude in Fig.~\ref{fig1:diag1}(b) into several parts which correspond to the muon
ground state $(n=2)$ and muon excited intermediate states $(n\not
=2)$. In this way first contribution with $n=2S$ takes the form:
\be
\label{eq:ver2s}
\Delta
\nu^{hfs}_{vert,2S}=\frac{8\alpha^2(1+\kappa_\mu)}{3\pi^2m_em_\mu}\int_0^\infty
k g(k^2)dk\int d{\bf x}_1\int d{\bf
x}_3\psi_{e}({\bf x}_3)
\Delta V_1(k,{\bf x}_3)G_e({\bf x}_1,{\bf x}_3)\Delta
V_\mu({\bf x}_1)\psi_{e}({\bf x}_1),
\ee
where $\Delta V_\mu({\bf x}_1)$ is defined by~\eqref{eq:deltavmu}  and
\be
\label{eq:deltav1}
\Delta V_1(k,{\bf x}_3)=\int d{\bf x}_4\psi_{\mu}({\bf
x}_4)\frac{\sin(k|{\bf x}_3-{\bf x}_4|)}{|{\bf x}_3-{\bf
x}_4|}\psi_{\mu}({\bf x}_4)=
\frac{\sin(k x_3)}{x_3}\frac{(1-3\frac{k^2}{W_\mu^2}+2\frac{k^4}{W_\mu^4})}
{\left[1+\frac{k^2}{W_\mu^2}\right]^4}.
\ee
Substituting the electron Green's function~\eqref{eq:green} in~\eqref{eq:ver2s} we
perform all coordinate integrations analytically and write \eqref{eq:ver2s}
in integral form ($a_1=k m_e/W_\mu$):
\be
\label{eq:ver2s1}
\Delta\nu^{hfs}_{vert,2S}=\nu_F\frac{\alpha(1+\kappa_\mu)}{2\pi^2}\left(\frac{m_e}
{\alpha M_\mu}\right)^2\left(\frac{M_e}{M_\mu}\right)\int_0^\infty
\frac{k g(k^2)a_1^4(1-3a_1^2k^2+2a_1^4k^4)dk}{(1+a_1^2)^8(a_1^2+M_e^2/M_\mu^2)^3}
\times
\ee
\bdis
\Bigl[a_1(2a_1^6+8a_1^5+8a_1^2+14)+\frac{M_e}{M_\mu}\Bigl(-2a_1^7-4a_1^5+14a_1+
14(a_1^2-1)^4\arctan\frac{1}{a_1}\Bigr)\Bigr]=0.067~MHz,
\edis
where we expand also the integral function in the numerator in small
parameter $M_e/M_\mu$ and carry out last integration numerically.

Second part of the vertex contribution (Fig.\ref{fig1:diag1}(b)) with $n=2P$
has the following general form:
\be
\label{eq:vert2p}
\Delta \nu^{hfs}_{vert,2P}=
\frac{8\alpha^2(1+\kappa_\mu)}{3\pi^2m_em_\mu}\int_0^\infty kg(k^2)dk \int d{\bf x}_1
\int d{\bf x}_2\int d{\bf x}_3\int d{\bf x}_4\psi^\ast_\mu({\bf x}_4)\psi^\ast_e({\bf x}_3)
\times
\ee
\bdis
\sum_{m=-1}^1
\psi_{\mu,2P,m}({\bf x}_4)\psi^\ast_{\mu,2P,m}({\bf x}_2)
\frac{\sin k|{\bf x}_4-{\bf x}_3|}{|{\bf x}_4-{\bf x}_3|}
\sum_{n\not=0}\frac{\psi_{e,n}({\bf x}_3)\psi^\ast_{e,n}({\bf x}_1)}
{E_{e0}-E_{e,n}}\Delta H({\bf x}_1,{\bf x}_2)\psi_\mu({\bf x}_2)\psi_e({\bf x}_1).
\edis
Then we make two subsequent integrations over coordinates ${\bf x}_2$ and
${\bf x}_4$:
\be
\label{eq:x2}
\delta\tilde V_2({\bf x}_1)=\int d{\bf x}_2 x_2 e^{-W_\mu x_2}\left(1-\frac{W_\mu x_2}{2}\right)
\frac{\alpha({\bf n}_2{\bf n}_4)}{|{\bf x}_2-{\bf x}_1|}=
\ee
\bdis
=\frac{2\pi\alpha}{W_\mu^3}\frac{e^{-x_1}({\bf n}_1{\bf n}_4)}{x^2_1}\left[x_1^4+4x_1^3+12x_1^2+24x_1-24e^{x_1}+24\right].
\edis
\be
\label{eq:x4}
\delta\tilde V_1({\bf x}_3)=\int d{\bf x}_4 x_4 e^{-W_\mu x_4}\left(1-\frac{W_\mu x_4}{2}\right)
({\bf n}_1{\bf n}_4)\frac{\sin k|{\bf x}_4-{\bf x}_3|}{|{\bf x}_4-{\bf x}_3|}=
\ee
\bdis
=\frac{48\pi}{W_\mu^3}({\bf n}_1{\bf n}_3)\frac{(kx_3\cos kx_3-\sin kx_3)\left(1-\frac{k^2}{W_\mu^2}\right)}
{x_3^2\left(1+\frac{k^2}{W_\mu^2}\right)^4}.
\edis
Finally, after analytical integration over coordinates ${\bf x}_1$ and ${\bf x}_3$ with the Green's
function~\eqref{eq:g11} we present the correction~\eqref{eq:vert2p} in the form:
\be
\label{eq:vert2p1}
\Delta \nu^{hfs}_{vert,2P}=-\nu_F\frac{3\alpha(1+\kappa_\mu)M_em_e^2}{\pi^2M_\mu(\alpha M_\mu)^2}
\int_0^\infty\frac{kg(k^2)a_1^3(1-a_1^2)dk}{(a_1^2+1)^4\Bigl(1+\frac{M_e}{M_\mu}\Bigr)^5
\Bigl(a_1^2+\frac{M_e^2}{M_\mu^2}\Bigr)^2\Bigl[a_1^2+\Bigl(1+\frac{M_e}{M_\mu}\Bigr)^2\Bigr]^4}
\ee
\bdis
\times\Bigl\{12-12a_1^2+\frac{M_e}{M_\mu}[75a_1^8+215a_1^6+209a_1^4+9a_1^2+15\pi a_1(1+a_1^2)^4-
75(1+a_1^2)^4a_1\arctan a_1+156]
\Bigr\}=
\edis
\bdis
=-0.309~MHz.
\edis
The expansion of integral function in the numerator of~\eqref{eq:vert2p1} is used up to terms of
order $O(M_e/M_\mu)$. In order to evaluate the vertex contribution due to excited states of the
muon we replace as above in~\eqref{eq:free} exact Green's function of the electron by free Green's
function. Then our starting point for the calculation is related with the following expression:
\be
\label{eq:vertex}
\Delta\nu^{hfs}_{vert,n\not=2}=-\frac{8\alpha^2(1+\kappa_\mu)}{3\pi^2m_em_\mu}\int_0^\infty kg(k^2)dk
\int d{\bf x}_3\int d{\bf x}_4\psi^\ast_\mu({\bf x}_4)\psi^\ast_e({\bf x}_3)
\frac{\sin k|{\bf x}_4-{\bf x}_3|}{|{\bf x}_4-{\bf x}_3|}\times
\ee
\bdis
\int d{\bf x}_1\int d{\bf x}_2\sum_{n\not=2}\psi_{\mu,n}({\bf x}_4)\psi^\ast_{\mu,n}({\bf x}_2)
\frac{e^{-b|{\bf x}_3-{\bf x}_1|}}{|{\bf x}_3-{\bf x}_1||{\bf x}_2-{\bf x}_1|}
\psi_{\mu}({\bf x}_2)\psi_e({\bf x}_1).
\edis
The integration over ${\bf x}_1$ is done in~\eqref{eq:jj}. The term proportional to
$1/b$ does not contribute because of the orthogonality of wave functions.
The contribution of second term in square brackets in~\eqref{eq:jj} can be
transformed initially by means of completeness condition:
\be
\label{eq:vertex1}
\Delta\nu^{hfs}_{vert,n\not=2}=\frac{8\alpha^3(1+\kappa_\mu)M_e}{3\pi^2m_em_\mu}\frac{W_e^{3/2}}{\sqrt{\pi}}
\int_0^\infty kg(k^2)dk\int d{\bf x}_3\int d{\bf x}_4\psi^\ast_{\mu}({\bf x}_4)
\psi^\ast_e({\bf x}_3)\frac{\sin k|{\bf x}_4-{\bf x}_3|}{|{\bf x}_4-{\bf x}_3|}\times
\ee
\bdis
\int d{\bf x}_2\left[\delta({\bf x}_2-{\bf x}_3)-\psi_{\mu,2S}({\bf x}_3)\psi^\ast_{\mu,2S}({\bf x}_2)-
\sum_{m=-1}^1\psi_{\mu,2P,m}({\bf x}_3)\psi^\ast_{\mu,2P,m}({\bf x}_2)\right]|{\bf x}_3-{\bf x}_2|
\psi_\mu({\bf x}_2).
\edis
Then the integration over particle coordinates is performed analytically in
all terms in~\eqref{eq:vertex1}. Remaining  integration over $k$ is
carried out numerically. Omitting intermediate expressions for numerous integrals
which are calculated in the same way as in previous corrections, we present final
numerical result
\be
\label{eq:vertex2}
\Delta\nu^{hfs}_{vert,n\not=2}=0.992~MHz.
\ee
The electron vertex corrections investigated in this section have the order $\alpha^5$
in the hyperfine interval. Summary value of all obtained contributions in second
order PT is equal to 0.750~MHz. Summing this number with the correction~\eqref{eq:amm2}
we obtain the value 5.828 MHz. It differs by a significant value 0.582 MHz from the
result 5.246 MHz which was used previously by many authors for the
estimation of the electron anomalous magnetic moment contribution (see \cite{LM,Amusia,Chen}.

\begin{figure}[htbp]
\centering
\includegraphics[scale=0.7]{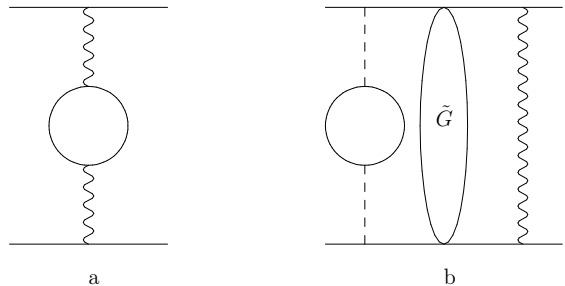}
\caption{Vacuum polarization effects. The dashed line represents the Coulomb photon.
The wave line represents the hyperfine part of the Breit potential. $\tilde G$ is
the reduced Coulomb Green's function.}
\end{figure}

\section{Effects of vacuum polarization}

The vacuum effects change significantly the interaction \eqref{eq:ham}-\eqref{eq:deltah} between
particles in muonic helium atom. One of the most important contributions is related
with the vacuum polarization (VP). The value of the electron vacuum polarization corrections to
the hyperfine splitting is determined by the parameter equal to the ratio of the Compton wave
length of the electron and the radius of the Bohr orbit in the subsystem $(\mu^4_2He)^+$:
$M_\mu\alpha/m_e$= $1.46727\ldots$. It is impossible to use the expansion in $\alpha$ for
such contributions. So, we calculate them performing the analytical and numerical integration
over the particle coordinates and other parameters. The effect of the electron vacuum
polarization leads to the appearance of a number of additional terms in the Hamiltonian
operator describing $(e-\alpha)$-, $(\mu-\alpha)$- and $(e-\mu)$-interactions
which we present in the form \cite{M3,M4,t4}:
\begin{equation}
\label{eq:vpealpha}
\Delta V_{VP}^{e\alpha}(x_e)=\frac{\alpha}{3\pi}\int_1^\infty d\xi\rho(\xi)\left(-\frac{2\alpha}{x_e}\right)
e^{-2m_e\xi x_e},~\rho(\xi)=\frac{\sqrt{\xi^2-1}(2\xi^2+1)}{\xi^4},
\end{equation}
\begin{equation}
\label{eq:vpmualpha}
\Delta V_{VP}^{\mu \alpha}(x_\mu)=\frac{\alpha}{3\pi}\int_1^\infty d\xi\rho(\xi)\left(-\frac{2\alpha}{x_\mu}\right)
e^{-2m_e\xi x_\mu},
\end{equation}
\begin{equation}
\label{eq:vpmue}
\Delta V_{VP}^{e\mu}(|{\bf x}_e-{\bf x}_\mu|)=\frac{\alpha}{3\pi}\int_1^\infty d\xi\rho(\xi)\frac{\alpha}{x_{e\mu}}
e^{-2m_e\xi x_{e\mu}},
\end{equation}
where $x_{e\mu}=|{\bf x}_e-{\bf x}_\mu|$. They give contributions in second order perturbation theory
and are discussed below. In first order perturbation theory the contribution of the vacuum
polarization is connected with the modification of the hyperfine splitting part of the Hamiltonian
\eqref{eq:hfsh} (the diagram in Fig.\ref{fig1:diag1}(a)). It is described by the expression
\cite{M1,M2}:
\begin{equation}
\label{eq:vp1g}
\Delta V_{VP}^{hfs}({\bf x}_{e\mu})=-\frac{8\pi\alpha(1+\kappa_\mu)}{3m_em_\mu}
\frac{{\mathstrut\bm\sigma}_1{\mathstrut\bm\sigma}_2}{4}\frac{\alpha}{3\pi}\int_1^\infty
\rho(\xi)d\xi\left[\pi\delta({\bf x_{e\mu}})-\frac{m_e^2\xi^2}{x_{e\mu}}e^{-2m_e\xi x_{e\mu}}
\right].
\end{equation}
Averaging the potential~\eqref{eq:vp1g} over the wave function~\eqref{eq:wave} we obtain the following
expression:
\begin{equation}
\label{eq:vp1g1}
\Delta\nu^{hfs,(1)}_{VP,e-\mu}=\frac{\alpha^2(1+\kappa_\mu)}{9m_em_\mu}\frac{W_e^3W_\mu^3}
{\pi^2}\int_1^\infty\rho(\xi)d\xi\int d{\bf
x}_e\int d{\bf x}_\mu e^{-W_\mu x_\mu}\left(1-\frac{W_\mu x_\mu}{2}\right)^2\times
\end{equation}
\begin{displaymath}
e^{-2W_e x_e}\left[\pi\delta({\bf x_\mu}-{\bf x}_e)-\frac{m_e^2\xi^2}{|{\bf x}_\mu-{\bf x}_e|}\right]
e^{-2m_e\xi|{\bf x}_\mu-{\bf x}_e|},
\end{displaymath}
where the superscript $(1)$ designates the correction of first order PT and the subscript $e-\mu$
corresponds to the electron-muon interaction.
There are two integrals over the muon and electron coordinates in~\eqref{eq:vp1g1}
which can be calculated analytically. As a result the correction~\eqref{eq:vp1g1} takes
the form of one-dimensional integral in the parameter $\xi$ ($a_2=m_e\xi/M_\mu\alpha$):
\begin{equation}
\label{eq:vp1g2}
\Delta\nu^{hfs,(1)}_{VP,e-\mu}=\nu_F\frac{\alpha(1+\kappa_\mu)
M_e}{6\pi M_\mu\left(1+\frac{M_e}{M_\mu}\right)^5}\int_1^\infty\frac{\rho(\xi)d\xi}
{(1+a_2)^4\left(a_2+\frac{M_e}{M_\mu}\right)^2}
\Biggl\{
a_2^4\left[1+\frac{M_e}{M_\mu}\left(4\frac{M_e}{M_\mu}-1\right)\right]+
\ee
\bdis
+2a_2^3\left(2+\frac{M_e}{M_\mu}\right)\left[1+\frac{M_e}{M_\mu}
\left(4\frac{M_e}{M_\mu}-1\right)\right]+a_2^2\left(1+\frac{M_e}{M_\mu}\right)\left[4+\frac{M_e}{M_\mu}
\left(11\frac{M_e}{M_\mu}-12\right)\right]+
\edis
\bdis
+4a_2\left[1+\frac{M_e}{M_\mu}-\frac{M^2_e}{M^2_\mu}+2\frac{M^3_e}{M^3_\mu}\right]+2\frac{M_e}{M_\mu}\left[1+
\frac{M_e}{M_\mu}\left(\frac{M_e}{M_\mu}-1\right)\right]
\Biggr\}=0.012~MHz.
\edis
In the leading order this contribution has fifth order in $\alpha$ and first order
in the electron-muon recoil parameter $M_e/M_\mu$. The similar contribution of the muon vacuum
polarization to the hyperfine splitting is extremely small ($\sim 10^{-6}$ MHz).

Let us consider corrections of the electron vacuum polarization
\eqref{eq:vpealpha}-\eqref{eq:vpmue} in the second order perturbation theory (the diagram
in Fig.~\ref{fig1:diag1}(b)). The contribution of the operator~\eqref{eq:vpealpha} to the hyperfine
splitting can be written as follows:
\begin{equation}
\label{eq:soptealpha}
\Delta\nu_{VP,e\alpha}^{hfs,(2)}=\frac{32\alpha^3}{9m_em_\mu}\int
d{\bf x}_1\int d{\bf x}_2\int d{\bf x}_3\int_1^\infty\rho(\xi)d\xi \psi^\ast_{\mu}({\bf
x}_3)\psi^\ast_{e}({\bf x}_3)\times
\end{equation}
\begin{displaymath}
{\sum_{n,m}}'\frac{\psi_{\mu,n}({\bf x}_3)\psi_{e,m}({\bf x}_3)\psi^\ast_{\mu,n}({\bf x}_2)
\psi^\ast_{e,m}({\bf x}_1)}{E_{\mu}+E_{e}-E_{\mu,n}-E_{e,m}}\frac{e^{-2m_e\xi x_1}}{x_1}\psi_{\mu}({\bf x}_2)
\psi_{e}({\bf x}_1).
\end{displaymath}
Here the summation is carried out over the complete system of the electron and muon eigenstates
excluding the state $1s^{(e)}_{1/2}2s^{(\mu)}_{1/2}$. Single contribution to the sum in~\eqref{eq:soptealpha}
is determined by the $2S$ muon intermediate state because of the wave function orthogonality condition:
\be
\label{eq:sopteal}
\Delta\nu_{VP,e\alpha}^{hfs,(2)}=\frac{32\alpha^3(1+\kappa_\mu)}{9m_em_\mu}
\int_1^\infty \rho(\xi)d\xi\int d{\bf x}_1\int d{\bf x}_3|\psi_\mu({\bf x}_3)|^2\psi_e({\bf x}_3)
G_e({\bf x}_1,{\bf x}_3)\frac{e^{-2m_e\xi x_1}}{x_1}\psi_{e}({\bf x}_1).
\ee
Using the electron Green's function~\eqref{eq:green} we perform
analytical integration over coordinates ${\bf x}_1$ and ${\bf x}_3$ and present~\eqref{eq:sopteal}
as follows:
\be
\label{eq:sopteal1}
\Delta\nu_{VP,e\alpha}^{hfs,(2)}=-\nu_F\frac{\alpha(1+\kappa_\mu)M_e}{3\pi M_\mu}\int_1^\infty
\rho(\xi)d\xi \frac{4+a_2(a_2+2)^2}{a_2(a_2+1)^4}=-0.028~MHz.
\ee

This contribution has the same order of the magnitude $O(\alpha^5M_e/M_\mu)$
as the previous one in first order perturbation theory. The similar calculation can be
performed in the case of muon-nucleus vacuum polarization operator~\eqref{eq:vpmualpha}. The intermediate
electron state is the 1S state and the reduced Coulomb Green's function of the system
transforms to the Green's function of the muon. The correction of the operator~\eqref{eq:vpmualpha}
to the hyperfine splitting is obtained in the following integral form:
\be
\label{eq:soptmualpha}
\Delta\nu^{hfs,(2)}_{VP,\mu
\alpha}=\frac{32\alpha^3(1+\kappa_\mu)}{9m_em_\mu}\int_1^\infty\rho(\xi)d\xi\int d{\bf x}_2\int d{\bf x}_3
\psi^\ast_\mu({\bf x}_3)\psi_\mu({\bf x}_2)|\psi_e({\bf x}_3)|^2
\frac{e^{-2m_e\xi x_2}}{x_2}G_\mu({\bf x}_2,{\bf x}_3),
\ee
where
\be
\label{eq:green2s}
G_\mu({\bf x}_2,{\bf x}_3)=-\frac{\alpha M_\mu^2}{16\pi}\frac{e^{-(x_2+x_3)/2}}{x_2 x_3}
\Bigl[8x_<-4x^2_<+8x_>+12x_<x_>-26x^2_<x_>+2x^3_<x_>-4x^2_>-
\ee
\bdis
-26x_<x^2_>+23x^2_<x^2_>
-x^3_<x^2_>+2x_<x^3_>-x^2_<x^3_>+4e^x(1-x_<)(x_>-2)x_>+4(x_<-2)x_<(x_>-2)x_>\times
\edis
\bdis
[-2C+Ei(x_<)-\ln(x_<)-\ln(x_>)].
\edis
Integrating over coordinates we obtain:
\be
\label{eq:soptmualpha1}
\Delta\nu^{hfs,(2)}_{VP,\mu
\alpha}=-\nu_F\frac{\alpha(1+\kappa_\mu)M_e}{2\pi M_\mu}\int_1^\infty\rho(\xi)d\xi
\frac{4a_2^4+8a_2^3+4a_2^2+6a_2+1}{(1+a_2)^6}=-0.099~MHz.
\ee

The vacuum polarization correction connected with the operator~\eqref{eq:vpmue}
in the second order perturbation theory is the most difficult for the calculation.
Indeed, in this case we have to consider the intermediate excited states both for the muon and
electron. We have divided total contribution into several parts. The first part in which the
intermediate muon is in the 2S state can be written as:
\be
\label{eq:soptmualpha2}
\Delta\nu^{hfs,(2)}_{VP,\mu e,2S}=
-\nu_F\frac{\alpha^3(1+\kappa_\mu)M_e^2}{3\pi W_\mu^2}\int_1^\infty\rho(\xi)d\xi
\int_0^\infty x_1^2 e^{-\frac{M_e}{M_\mu}x_1} dx_1\int_0^\infty x_3^2dx_3\left(1-\frac{x_3}{2}\right)^2
e^{-x_3(1+\frac{M_e}{M_\mu})}
\ee
\bdis
\times\Delta V_{VP,\mu}(x_1)\bigl[\frac{1}{\frac{M_e}{M_\mu}x_>}-\ln\frac{M_e}{M_\mu}x_>-
\ln\frac{M_e}{M_\mu}x_<+Ei(\frac{M_e}{M_\mu}x_<)+\frac{7}{2}-2C-\frac{M_e(x_1+x_3)}{2M_\mu}+\frac{1-e^{\frac{M_e}{M_\mu}x_<}}
{\frac{M_e}{M_\mu}x_<}\bigr],
\edis
where the function $V_{VP~\mu}(x_1)$ is equal to
\be
\label{eq:intj}
\Delta V_{VP,\mu}(x_1)=\int_0^\infty x_2^2 e^{-x_2}\left(1-\frac{x_2}{2}\right)^2dx_2 \int_1^1 dz
\frac{e^{-a_2|{\bf x}_1-{\bf x}_2|}}{|{\bf x}_1-{\bf x}_2|}
=
\ee
\bdis
=\frac{1}{2x_1(a^2_2-1)^4}\Bigl\{8e^{-a_2x_1}(2a_2^4+3a_2^2+1)
+e^{-x_1}\Bigl[a_2^6x_1(x_1-2)^2+a_2^4(x_1(10-3(x_1-2)x_1)-16)+
\edis
\bdis
+a_2^2(3x_1^3-8x_1-24)-x_1(x_1(x_1+2)+6)-8)\Bigr]
\Bigr\}.
\edis
Substituting \eqref{eq:intj} in \eqref{eq:soptmualpha2} we obtain the following result with an accuracy $O(M_e/M_\mu)$:
\be
\label{eq:soptmualpha2}
\Delta\nu^{hfs,(2)}_{VP,\mu e,2S}=-\nu_F\frac{\alpha M_e(1+\kappa_\mu)}{768\pi M_\mu}\int_1^\infty\frac{\rho(\xi)d\xi}
{a_2(1+a_2)^8}\Bigl(77a_2^7+616a_2^6+2151a_2^5+4272a_2^4+
\ee
\bdis
+5267a_2^3+4168a_2^2+1929a_2+512\Bigr)=
-0.009~MHz.
\edis
The second part of the vacuum polarization correction to the hyperfine splitting due to the $2P$ muon
state can be presented as
\begin{equation}
\label{eq:soptmue2p}
\Delta\nu^{hfs,(2)}_{VP,\mu e,2P}=\frac{16\alpha^3(1+\kappa_\mu)}{9m_em_\mu}\int d{\bf x}_1\int d{\bf x}_2
\int d{\bf x}_3\psi^\ast_\mu({\bf x}_3)\psi^\ast_e({\bf x}_3)
\sum_{m=-1}^1\psi_{\mu,2P,m}({\bf x}_3)\psi^\ast_{\mu,2P,m}({\bf x}_2)\times
\end{equation}
\begin{displaymath}
\times\sum_{n\not =0}\frac{\psi_{e,n}({\bf x}_3)\psi^\ast_{e,n}({\bf x}_1)}{E_e-E_{e,n}}
\int_1^\infty\rho(\xi)d\xi\frac{e^{-2m_e\xi|{\bf x}_2-{\bf x}_1|}}{|{\bf x}_2-{\bf x}_1|}
\psi_{\mu}({\bf x}_2)\psi_{e}({\bf x}_1).
\end{displaymath}
Averaging over the directions of the vector ${\bf n}_3$ $<({\bf n}_3{\bf n}_2)({\bf n}_3
{\bf n}_1)>$=$({\bf n}_1{\bf n}_2)/3$ and evaluating the integral
\be
\label{eq:inttj}
\Delta \tilde V_{VP,\mu}(x_1)=\int_0^\infty x_2^3 e^{-x_2}\left(1-\frac{x_2}{2}\right)dx_2 \int_1^1 zdz
\frac{e^{-a_2|{\bf x}_1-{\bf x}_2|}}{|{\bf x}_1-{\bf x}_2|}
=\frac{1}{x_1^2(a^2_2-1)^4}\Bigl\{e^{-x_1}\times
\ee
\bdis
\Bigl[24(a_2^2+1)(x_1+1)-12x_1^2(a_2^4-1)+
2x_1^3(a_2^6-3a_2^2+2)-x_1^4(a_2^2-1)^3\Bigr]-24e^{-a_2x_1}(a_2^2+1)(a_2x_1+1)\Bigr\},
\edis
we find this contribution numerically:
\be
\label{eq:soptmue2p1}
\Delta\nu^{hfs,(2)}_{VP,\mu e,2P}=-\nu_F\frac{9\alpha(1+\kappa_\mu)M_e}{192\pi M_\mu}\int_1^\infty
\frac{\rho(\xi)d\xi}{(1+a_2)^8}\Bigl[5a_2^6+40a_2^5+139a_2^4+272a_2^3+
\ee
\bdis
+323a_2^2+232a_2+93\Bigr]=-0.019~MHz.
\edis
Finally, there exists the contribution of the muon intermediate states with
$n\not=2S,2P$:
\be
\label{eq:nnot}
\Delta\nu^{hfs,(2)}_{n\not=2S,2P}=-\frac{16\alpha^3(1+\kappa_\mu)}{9m_em_\mu}\int d{\bf x}_1
\int d{\bf x}_2\int d{\bf x}_3\psi^\ast_\mu({\bf x}_3)\psi_e^\ast({\bf x}_3)\sum_{n\not=2}
\psi_{\mu,n}({\bf x}_3)\psi^\ast_{\mu,n}({\bf x}_2)\times
\ee
\bdis
G_e({\bf x}_1,{\bf x}_3,E_\mu+E_e-E_{\mu,n})\int_1^\infty\rho(\xi)d\xi
\frac{e^{-2m_e\xi|{\bf x}_2-{\bf x}_1|}}{|{\bf x}_2-{\bf x}_1|}\psi_\mu({\bf x}_2)\psi_e({\bf x}_1).
\edis
Then we have replaced in~\eqref{eq:nnot} exact electron Coulomb
Green's function $G_e$ by free electron Green's function~\eqref{eq:free} which
contains the parameter $b=[2M_e(E_{\mu,n}-E_{\mu}-E_{e}]^{1/2}$. We also
replace the electron wave functions by their values at the origin
neglecting higher order recoil corrections. After
that the integration over ${\bf x}_1$ can be done analytically:
\be
\label{eq:i2}
I_2=\int d{\bf x}_1\frac{e^{-b|{\bf x}_3-{\bf x}_1|}}{|{\bf x}_3-{\bf x}_1|}
\frac{e^{-2m_e\xi|{\bf x}_2-{\bf x}_1|}}{|{\bf x}_2-{\bf x}_1|}=-\frac{4\pi}{W_\mu}
\frac{e^{-b_1|{\bf x}_3-{\bf x}_2|}-e^{-a_2|{\bf x}_3-{\bf x}_2|}}
{|{\bf x}_3-{\bf x}_2|(b_1^2-a_2^2)}=
\ee
\bdis
=-\frac{4\pi}{W_\mu}\Bigl[\frac{1-e^{-a_2|{\bf x}_3-{\bf x}_2|}}{a_2^2|{\bf x}_3-{\bf x}_2|}-
\frac{2b_1}{a_2^2}\ldots\Bigr]
\edis
where we have performed small-parameter $b_1=b/W_\mu$ expansion. For the further transformation
we use the completeness relation~\eqref{eq:delta} dividing total correction~\eqref{eq:nnot} into
three terms. The coordinate integration in each of the terms can be presented individually in
analytical form. As a result the correction~\eqref{eq:nnot} is reduced to one-dimensional integral
over the parameter $\xi$ which is evaluated with high accuracy numerically:
\be
\label{eq:nnot1}
\Delta\nu^{hfs,(2)}_{n\not=2S,2P}=\nu_F\frac{\alpha(1+\kappa_\mu)M_e}{384\pi M_\mu}\int_1^\infty
\frac{\rho(\xi)d\xi}{(1+a_2)^7}\Bigl(256a_2^6+1731a_2^5+4949 a_2^4+
\ee
\bdis
+7686a_2^3+
6874a_2^2+3375a_2+665\Bigr)=0.076~MHz.
\edis
We preserve in~\eqref{eq:nnot1} only leading order term in the ratio $M_e/M_\mu$. In whole,
in~\eqref{eq:nnot1} we have second order correction in two small parameters $\alpha$ and
$M_e/M_\mu$. We keep in~\eqref{eq:i2} only first term in the square brackets because other
terms give recoil corrections of higher order. Numerical values of obtained corrections are
presented in Table I.

\begin{figure}[htbp]
\centering
\includegraphics[scale=0.8]{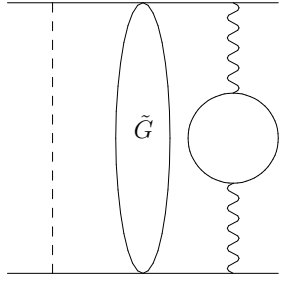}
\caption{Vacuum polarization effects in second order perturbation theory. The dashed line represents
first part of the potential
$\Delta H$ (3). The wave line represents the hyperfine part of the Breit potential.}
\label{fig2:diag2}
\end{figure}

There exists another contribution of the second order perturbation
theory in which we have the vacuum polarization perturbation
connected with the hyperfine splitting part of the Breit potential
\eqref{eq:ham} (see Fig.~\ref{fig3:diag3}). Other perturbation is determined by $\Delta H$
\eqref{eq:deltah}. We can divide this correction into three parts. One part corresponds to
the muon in the $2S$-state. The $\delta$-function term in~\eqref{eq:vp1g} gives the following
contribution at $n=2$ (compare with~\eqref{eq:nu_sopt_2s}):
\be
\label{eq:vp11}
\Delta\nu^{hfs,(2)}_{VP,11,2S}=\nu_F(1+\kappa_\mu)\frac{\alpha}{3\pi}\int_1^\infty\rho(\xi)d\xi
\frac{631M_e}{256 M_\mu}.
\ee
Obviously this integral is divergent. So, we have to consider the contribution of second term in
\eqref{eq:vp1g} to the hyperfine splitting which is determined by more complicated
expression:
\be
\label{eq:vp12}
\Delta\nu^{hfs,(2)}_{VP,12,2S}=-\frac{16\alpha^2(1+\kappa_\mu)m_e^2}{9\pi m_em_\mu}\int_1^\infty
\rho(\xi)\xi^2d\xi \int d{\bf x}_4|\psi_{\mu}({\bf x}_4)|^2\int d{\bf x}_3\psi_{e}({\bf x}_3)
\frac{e^{-2m_e|{\bf x}_3-{\bf x}_4|}}{|{\bf x}_3-{\bf x}_4|}\times
\ee
\bdis
\times\sum_{n'\not =0}
\frac{\psi_{e,n'}({\bf x}_3)\psi^\ast_{e,n'}({\bf x}_1)}{E_{e}-E_{en'}}\Delta V_\mu({\bf x}_1)
\psi_{e}({\bf x}_1),
\edis
where the matrix elements entering in~\eqref{eq:vp12} are defined by~\eqref{eq:deltavmu} and \eqref{eq:intj}.
Integrating over all coordinates in~\eqref{eq:vp12}  and summing it with~\eqref{eq:vp11} we obtain
the following convergent integral in $\xi$ in the leading order with respect to the ratio
$(M_e/M_\mu)$:
\be
\label{eq:vp1112}
\Delta\nu^{hfs,(2)}_{VP,11,2S}+\Delta\nu^{hfs,(2)}_{VP,12,2S}=
\nu_F\frac{\alpha(1+\kappa_\mu)M_e}{3\pi M_\mu}\int_1^\infty\frac{\rho(\xi)d\xi}{(a_2+1)^8}
\Bigl(51a_2^6+408a_2^5+
\ee
\bdis
+1177a_2^4+1872a_2^3+2029a_2^2+1464a_2+631\Bigr)=0.004~MHz.
\edis

\begin{figure}[htbp]
\centering
\includegraphics[scale=0.8]{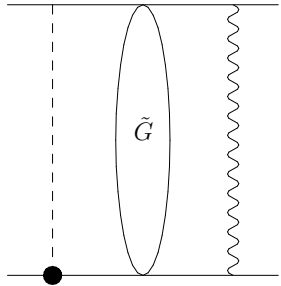}
\caption{Nuclear structure effects in second order perturbation theory. The bold
point represents the nuclear vertex operator.
The wave line represents the hyperfine part of the Breit potential.}
\label{fig3:diag3}
\end{figure}

Next correction arises from the muon in $2P$ intermediate state. Making the same division of
total contribution into two parts as in~\eqref{eq:vp1112} we present here final result:
\be
\label{eq:vp1112p}
\Delta\nu^{hfs,(2)}_{VP,11,2P}+\Delta\nu^{hfs,(2)}_{VP,12,2P}=
-\nu_F\frac{3\alpha(1+\kappa_\mu)M_e}{256\pi M_\mu}\int_1^\infty\frac{\rho(\xi)d\xi}{(a_2+1)^8}
\Bigl[5a_2^6+40a_2^5+139a_2^4+
\ee
\bdis
+272a_2^3+323a_2^2+232a_2+93\Bigr]=-0.005~MHz.
\edis

Let us consider now the terms with $n\not =2S,2P$:
\be
\label{eq:3vp}
\Delta\nu^{hfs,(2)}_{VP,\mu-e,n\not= 2S,2P}=-\frac{16\alpha^3(1+\kappa_\mu)M_e}{18\pi^2m_em_\mu}\int_1^\infty
\rho(\xi)d\xi\int d{\bf x}_1\int d{\bf x}_2\int d{\bf x}_3\int d{\bf x}_4\psi^\ast_\mu({\bf x}_4)
\psi^\ast_e({\bf x}_3)\times
\ee
\bdis
\Bigl[\pi\delta({\bf x}_3-{\bf x}_4)-\frac{m_e^2\xi^2}{|{\bf x}_3-{\bf x}_4|}e^{-2m_e\xi|{\bf x}_3-{\bf x}_4|}\Bigr]
\sum_{n\not=2S,2P}\psi_{n,\mu}({\bf x}_4)\psi^\ast_{\mu,n}({\bf x}_2)\frac{e^{-b|{\bf x}_3-{\bf x}_1|}}
{|{\bf x}_3-{\bf x}_1||{\bf x}_2-{\bf x}_1|}\psi_\mu({\bf x}_2)\psi_e({\bf x}_1),
\edis
where free Green's function of the electron~\eqref{eq:free} is used.
Changing $\psi_e({\bf x}_1)\to\psi_e(0)$ and using first nonzero term in the expansion~\eqref{eq:jj}
we present the contribution of the $\delta$-function from~\eqref{eq:3vp} in the form:
\be
\label{eq:31}
\Delta\nu^{hfs,(2)}_{VP,11,n\not=2S,2P}=-\nu_F\frac{\alpha(1+\kappa_\mu)M_e}{3\pi M_\mu}\int_1^\infty
\rho(\xi)d\xi\frac{665}{128}.
\end{equation}
As expected the integral~\eqref{eq:31} in $\xi$ is divergent. We should consider together with it
another term from \eqref{eq:3vp}:
\be
\label{eq:32}
\Delta\nu^{hfs,(2)}_{VP,12,n\not 2S,2P}=-\nu_F\frac{2\alpha^2(1+\kappa_\mu) M_e m_e^2}
{3\pi^2}\int_1^\infty\rho(\xi)d\xi\int d{\bf x}_2\int d{\bf x}_3\int d{\bf x}_4
\frac{e^{-b|{\bf x}_3-{\bf x}_4|}}{|{\bf x}_3-{\bf x}_4|}\psi_\mu({\bf x}_4)\times
\ee
\bdis
\left[\delta({\bf x}_2-{\bf x}_4)-\psi_\mu({\bf x}_2)\psi^\ast_\mu({\bf x}_4)-
\sum_{m=-1}^1\psi_{\mu,2P,m}({\bf x}_2)\psi^\ast_{\mu,2P,m}({\bf x}_4)\right]|{\bf x}_3-
{\bf x}_2|\psi_\mu({\bf x}_2).
\edis
All three terms in square brackets give the contributions in~\eqref{eq:32}. They can be
evaluated in the same way as in previous sections. The sum of expressions~\eqref{eq:31} and \eqref{eq:32}
is determined by finite integral which is calculated numerically:
\be
\label{eq:3132}
\Delta\nu^{hfs,(2)}_{VP,11,n\not=2S,2P}+\Delta\nu^{hfs,(2)}_{VP,12,n\not 2S,2P}=
\nu_F\frac{\alpha(1+\kappa_\mu)M_e}{3\pi M_\mu}\int_1^\infty\frac{\rho(\xi)d\xi}
{128a_2(a_2+1)^9}\Bigl(-5a_2^9+16a_2^8+
\ee
\bdis
+403a_2^7+2075a_2^6+4596a_2^5+7310a_2^4+6559a_2^3+4511a_2^2+1639a_2+256\Bigr)
=0.012~MHz.
\edis
All vacuum polarization contributions to the hyperfine splitting are presented in Table I.
In their calculation we use systematically the expansion in $M_e/M_\mu$ retaining first order
terms. So, all corrections evaluated in this section are of second order in two small
parameters $\alpha$ and $M_e/M_\mu$.

\begin{table}
\caption{\label{t1}Hyperfine singlet-triplet splitting of the excited state in
muonic helium atom.}
\bigskip
\begin{ruledtabular}
\begin{tabular}{|c|c|c|}  \hline
Contribution to the HFS & $\Delta\nu^{hfs}$, MHz & Reference   \\   \hline
The Fermi splitting & 4516.915  & (5)  \\  \hline
The muon anomalous magnetic &  5.266   &  (5)  \\
moment correction  $a_\mu\nu_F$                &      &    \\    \hline
Recoil corrections in $M_e/M_\mu$ & -132.608 &   (5)  \\
in first order PT   &       &      \\    \hline
Recoil correction in $M_e/M_\mu$ in second &   61.127   &  (10)  \\
order PT, muon $2S$ state   &   &        \\     \hline
Recoil correction in $M_e/M_\mu$ in second&   -70.919   &  (13)  \\
order PT, muon $2P$ state   &   &        \\     \hline
Recoil correction in $M_e/M_\mu$ in second &   -109.978   &  (18), (21)  \\
order PT, muon excited states   &   &        \\     \hline
Recoil correction of order $\frac{M_e}{m_\alpha}\sqrt{\frac{M_e}{M_\mu}}\alpha^4$ in second& 0.189 & (27)   \\
order PT, muon excited states &     &      \\    \hline
Recoil correction of order $\frac{M^2_e}{m_\alpha M_\mu}\alpha^4$ in second & 0.074 & (29)   \\
order PT, muon excited states &     &      \\    \hline
Recoil correction of order & 0.812  &  (31), \cite{Chen}  \\
$\alpha^5(m_e/m_\mu)\ln(m_e/m_\mu)$    &     &    \\   \hline
Recoil correction $\frac{69m^2_e}{2m^2_\mu}\ln\frac{m_\mu}{m_e}\nu_F$ & 19.434 & \cite{Amusia}   \\
in third order PT &     &      \\    \hline
Electron vertex correction of order $\alpha^5$ & 5.078   &  (33)   \\
in first order PT   &       &            \\      \hline
Electron vertex correction of order $\alpha^5$ &  0.750  &  (36), (40), (43)   \\
in second order PT   &       &            \\      \hline
One-loop VP contribution in $1\gamma$ &  0.012 &  (49)     \\
$(e\mu)$ interaction of order $\alpha^5$   &  &     \\   \hline
One-loop VP contribution in the & 0.028 &   (52)   \\
electron-nucleus interaction in the&         &            \\
second order PT of order $\alpha^5$   &     &          \\   \hline
One-loop VP contribution in the &  0.099 &   (55)   \\
muon-nucleus interaction in the&        &            \\
second order PT of order $\alpha^5$   &     &          \\   \hline
One-loop VP contribution in the & 0.059  &  (58),(61),(64),(67),     \\
electron-muon interaction in the&        &   (68),(72)         \\
second order PT of order $\alpha^5$   &     &          \\   \hline
Nuclear structure correction of order  & -0.015 & (75),(76) \\
$\alpha^6$ in second order PT &       &        \\   \hline
Relativistic correction of order $\alpha^6$  & -0.050  & \cite{Chen} \\ \hline
Electron vertex
correction of order $\alpha^6$ &  -0.615 &  \cite{BE,KP,KKS,EGS}  \\
\hline Summary contribution & $4295.658$  &      \\  \hline
\end{tabular}
\end{ruledtabular}
\end{table}

\section{Nuclear structure corrections}

In the leading order in $\alpha$ the nuclear structure corrections
to hyperfine splitting are determined by the charge radius of $\alpha$-particle $r_\alpha$.
Considering the interaction between the muon and the nucleus we can present
the nuclear structure correction in the interaction operator in the
form \cite{EGS}:
\be
\label{eq:nsc}
\Delta V_{str,\mu-\alpha}({\bf r}_\mu)=\frac{2}{3}\pi Z\alpha<r^2_\alpha>\delta({\bf r}_\mu),
\ee
where the subscript str designates the structure correction.
The contribution of the operator $\Delta V_{str,\mu-\alpha}$ to the hyperfine splitting appears
in second order perturbation theory (see the diagram in Fig.\ref{fig3:diag3}).
We can write it in the integral form:
\be
\label{eq:str1}
\Delta\nu_{str,\mu-\alpha}^{hfs}=\frac{64\pi^2\alpha^2(1+\kappa_\mu)r_\alpha^2}{9m_em_\mu}
\frac{W_\mu^{3/2}}{2\sqrt{2\pi}}\int d{\bf x}_3\psi^\ast_{\mu}({\bf x}_3)|\psi_{e}({\bf x}_3)|^2
G_\mu({\bf x}_3,0,E_{\mu}).
\ee
Using the muon Green's function with one zero argument we make analytical
integration in~\eqref{eq:str1} and present final result in the form:
\be
\label{eq:str2}
\Delta\nu_{str,\mu-\alpha}^{hfs}=-\nu_F\frac{8\alpha^2M_\mu^2
r_\alpha^2(1+\kappa_\mu)}{3}\left(6\frac{M_e}{M_\mu}-41\frac{M_e^2}{M_\mu^2}
+\ldots\right)=-0.014~MHz.
\ee
Numerical value is obtained by means of the charge radius of the
$\alpha$-particle $r_\alpha=1.676$~fm. The same approach can be used
in the study of the electron-nucleus interaction. The electron feels
as well the distribution of the electric charge of $\alpha$
particle. The corresponding contribution of the nuclear structure
effect to the hyperfine splitting is the following:
\be
\label{eq:str3}
\Delta\nu_{str,e-\alpha}^{hfs}=-\nu_F\frac{4\alpha^2
M_\mu^2r_\alpha^2}{3}\left[\frac{1}{2}\frac{M_e}{M_\mu}-\frac{M_e^2}{M_\mu^2}\left(\frac{3}{2}+
\ln\frac{M_e}{M_\mu}\right)+\ldots\right]
=-0.001~MHz.
\ee
We have included in Table I the total nuclear structure contribution
which is equal to the sum of numerical values~\eqref{eq:str2} and \eqref{eq:str3}.

\section{Conclusion}

In this work we have performed the analytical and numerical calculation of some
basic contributions to the hyperfine splitting of the excited state $1s^{(e)}_{1/2}2s^{(\mu)}_{1/2}$
in muonic helium atom connected with the recoil effects, electron vertex
corrections, vacuum polarization and nuclear structure effects. We used the
perturbation theory for this task as it was formulated previously in~\cite{LM}.
As was shown by Lakdawala and Mohr that the ground state hyperfine splitting in
muonic helium atom can be expressed analytically by a series in the ratio $M_e/M_\mu$.
We extend their method to the case of the excited state $1s^{(e)}_{1/2}2s^{(\mu)}_{1/2}$
accounting for not only purely recoil corrections but also other vacuum effects controlled
by usual small parameter $\alpha$ in QED. Our results are presented in Table I in which we
give also references to the calculations of some
other corrections which are not considered here. We have included in the total value of the hyperfine
splitting the recoil logarithmic correction obtained in~\cite{Amusia} in third order PT, the relativistic
correction obtained in~\cite{Chen}, the electron vertex correction of order $O(\alpha^2)$
known from the theory of hydrogenic atoms \cite{EGS,BE,KP,KKS}. Our numerical result 4295.66~MHz is,
in whole, in the agreement with the earlier performed calculations by the perturbation theory and
variational method:
4291.50~MHz \cite{Amusia}, $4287.01\pm 0.10$~MHz \cite{Chen}.
Our analytical expressions of different recoil corrections are in agreement with the
calculation in~\cite{Amusia}. Comparing our results \eqref{eq:nu_sopt_2s}, \eqref{eq:nu_sopt_2pa},
\eqref{eq:2S2P}, \eqref{eq:s12}, \eqref{eq:malpha3} with~\cite{Amusia} we can find their coincidence
in the leading order in $m_e/m_\mu$, $m_e/m_\alpha$ and logarithmic terms $\frac{m_e^2}{m_\mu^2}\ln\frac{m_\mu}{m_e}$.
We should take into account that the used definition of reduced masses $M_e$ and $M_\mu$
of the particles is different from $m_e^\ast$ and $m_\mu^\ast$ in~\cite{Amusia}. The correspondence between our
results and basic expression (62) from~\cite{Amusia} can be revealed if we use the
relation: $M_e/M_\mu=m_e^\ast/m_\mu+m_e^\ast/m_\alpha$. It is interesting to note that the
recoil correction of order $m_e/m_\alpha$ appears in our kinematics from the
definitions of reduced masses $M_e$ and $M_\mu$. The recoil Hamiltonian $\Delta H_{rec}$ gives
the contribution in second order perturbation theory of order $M_e^2/m_\alpha M_\mu$ \eqref{eq:malpha3}.
Numerical differences of our results from~\cite{Amusia} are determined by
different choice of three-particle kinematics,
higher order corrections $M_e^2/M_\mu^2$ which we account for in intermediate expressions
and calculated vacuum polarization effects of order $\alpha M_e/M_\mu$. So, for example, numerical
value of the recoil correction of the leading order $\left(-6\frac{M_e}{M_\mu}\right)\nu_F$ in~\eqref{eq:ef}
is equal to $-134.769$~MHz. This number differs by more than 2~MHz from the value $-132.608$~MHz
presented in Table~I because we preserve corrections of higher order presenting \eqref{eq:ef} in the form $\nu_F(1+\kappa_\mu)(1-6M_e/M_\mu+21M_e^2/M_\mu^2-55M_e^3/M_\mu^3)$.
Another important source of the difference is specified by more accurate calculation of the
electron vertex correction in section III related with the Pauli form factor. We use exact expression for it
making numerical integration in corresponding amplitudes. Finally, we use in hyperfine splitting
part of the Hamiltonian the factor $(1+\kappa_\mu)$ related with the muon anomalous magnetic moment
what also gives definite modifications of numerical results. Our summary value of the vacuum polarization
effects 0.198~MHz agrees with the result 0.229~MHz obtained by variational method calculation in~\cite{Chen}.

Effects of the nuclear structure give negligibly small contribution to hyperfine
splitting. In muonic helium atom the muon strongly interacts with $\alpha$-particle
at distances of order $r\sim 1/M_\mu\alpha$. It was discovered recently in the
muonic hydrogen Lamb shift experiment that the experimental value
differs from theoretical one on 0.3~meV with CODATA value of the proton
charge radius \cite{nature}. This disagreement between the proton charge radius extracted from Lamb
shift measurements of muonic and electronic hydrogen can be considered as the appearance
of new muon physics \cite{carlson}. The plans to measure the $(2S-2P)$ transition frequencies in muonic
helium ions $(\mu^4_2He)^+$ and $(\mu^3_2He)^+$ with an accuracy of 50~ppm can help to solve the
proton radius puzzle \cite{nez}. In the case of muonium the most precise comparison between theory
and experiment is possible \cite{EGS,eides}. It will be interesting to use muonic helium atom for the
comparison of theoretical value of hyperfine splitting 4295.66~MHz with experimental data for
the revealing of new interactions in muonic sector.
Main theoretical error in our calculation is connected with uncalculated
in full recoil contribution  of second order $\nu_F\frac{M_e^2}{M_\mu^2}
\approx 0.11$~MHz. In addition, the electron vertex corrections in two-photon
processes which are estimated in this work very approximately also give essential error
$\nu_F\alpha^2\approx 0.25$~MHz. Thereby, the total theoretical error does not exceed
$\pm 0.40$~MHz, but large values of coefficients can essentially change
preliminary estimate. Further improvements of the theoretical result require more careful
construction of the three-particle interaction operator connected with the multiphoton
exchanges.

\begin{acknowledgments}
This work is supported by the Russian Foundation for Basic Research
(grant No.11-02-00019) and the Ministry of Education and
Science of Russian Federation (government order for Samara State U. No. 2.870.2011).
\end{acknowledgments}

\end{document}